\documentclass[letterpaper,twocolumn,10pt]{article}
\usepackage{zhanggroup}

\usepackage{amsfonts}
\usepackage{booktabs}
\usepackage{caption}
\usepackage{cite}
\usepackage{etoolbox}
\usepackage{float}
\usepackage{multirow}
\usepackage{subcaption}
\usepackage{tikz}
\usepackage{url}
\usepackage{xspace}
\usepackage[pagebackref=true,breaklinks=true,colorlinks,bookmarks=false]{hyperref}

\newcommand*\emptycirc[1][1ex]{\tikz\draw (0,0) circle (#1);}
\newcommand*\halfcirc[1][1ex]{%
	\begin{tikzpicture}
	\draw[fill] (0,0)-- (90:#1) arc (90:270:#1) -- cycle ;
	\draw (0,0) circle (#1);
	\end{tikzpicture}}
\newcommand*\fullcirc[1][1ex]{\tikz\fill (0,0) circle (#1);} 
\newcommand{\mypara}[1]{\noindent{\bf {#1}.}}

\begin{document}

\date{}

\title{\bf Backdoor Attacks in the Supply Chain of Masked Image Modeling}

\author{
Xinyue Shen\textsuperscript{1}\ \ \
Xinlei He\textsuperscript{1}\ \ \
Zheng Li\textsuperscript{1}\ \ \
Yun Shen\textsuperscript{2}\ \ \
Michael Backes\textsuperscript{1}\ \ \
Yang Zhang\textsuperscript{1}\ \ \
\\
\\
\textsuperscript{1}\textit{CISPA Helmholtz Center for Information Security} \ \ \ 
\textsuperscript{2}\textit{NetApp}
}

\maketitle

\begin{abstract}
Masked image modeling (MIM) revolutionizes self-supervised learning (SSL) for image pre-training. 
In contrast to previous dominating self-supervised methods, i.e., contrastive learning, MIM attains state-of-the-art performance by masking and reconstructing random patches of the input image. 
However, the associated security and privacy risks of this novel generative method are unexplored. 
In this paper, we perform the first security risk quantification of MIM through the lens of backdoor attacks. 
Different from previous work, we are the first to systematically threat modeling on SSL in every phase of the model supply chain, i.e., pre-training, release, and downstream phases. 
Our evaluation shows that models built with MIM are vulnerable to existing backdoor attacks in release and downstream phases and are compromised by our proposed method in pre-training phase.
For instance, on CIFAR10, the attack success rate can reach 99.62\%, 96.48\%, and 98.89\% in the downstream phase, release phase, and pre-training phase, respectively.
We also take the first step to investigate the success factors of backdoor attacks in the pre-training phase and find the trigger number and trigger pattern play key roles in the success of backdoor attacks while trigger location has only tiny effects. 
In the end, our empirical study of the defense mechanisms across three detection-level on model supply chain phases indicates that different defenses are suitable for backdoor attacks in different phases. 
However, backdoor attacks in the release phase cannot be detected by all three detection-level methods, calling for more effective defenses in future research.
\end{abstract}

\section{Introduction}
\label{section:introduction}

The self-supervised pre-training task has been dominant by contrastive learning, a discriminative method, in the computer vision domain since 2018~\cite{ZZSYZK22}.
Recently, with the advent of the Transformer architecture, masked image modeling (MIM), a generative method, has successfully surpassed contrastive learning and reached state-of-the-art performance on self-supervised pre-training tasks~\cite{BDW21, HCXLDG21, CDWXMWHLZW22, XZCLBYDH21}.
Compared with contrastive learning which aims to align different augmented views of the same image, MIM learns from predicting properties of masked patches from unmasked parts.
It plays as a milestone that bridges the gap between visual and linguistic self-supervised pre-training methods, and has quickly emerged variants in applications such as images~\cite{ABCGL22, BMAZ22}, video~\cite{WFXWYF21, TSWW22}, audio~\cite{BPH22}, and graph~\cite{TLHCCH22}.
However, as an iconic method settling in another branch of SSL, the associated security risks caused by the mask-and-predict mechanism and novel architectures of MIM are still unexplored.

\mypara{Our Contributions}
In this paper, we perform the first security risk quantification of MIM through the lens of backdoor attacks.
Different from previous work, we are the first to systematically categorize the threat models on MIM in every phase of model supply chain, i.e., pre-training, release, and downstream phases (see~\autoref{section:attack_taxonomy} for more details).
Our evaluation shows that models built with MIM are vulnerable to existing backdoor attacks in release and downstream phases.
For instance, in the downstream phase, with only 0.1\% poisoning rate (e.g., only 50 training samples on CIFAR10) and 0.05\% occupied area of the image, the attacker can achieve 89.37\% ASR on CIFAR10.

We also observe that previous attack~\cite{STKP21}, which successfully backdoors contrastive learning in the pre-training phase, cannot achieve satisfying attack performance on MIM.
The ASR is only 2.83\% and 13.78\% higher than the baseline on CIFAR10 and STL10, respectively.
To improve the attack performance in the pre-training phase, we propose a simple yet effective method: increasing the number of triggers in the span of the whole image.
We observe that, with our method, the ASR rises to 98.89\% and 97.74\% on CIFAR10 and STL10 datasets, respectively.

To further investigate the hardest yet rarely explored scenario, i.e., the pre-training phase, we conduct comprehensive ablation studies on the properties of triggers, i.e., pattern, location, number, size, and poisoning rate.
We find that trigger pattern and trigger number are key components that affect attack performance on MIM, which is different from a previous study on contrastive learning~\cite{STKP21}.
We utilize the white trigger and publicly released triggers of Hidden Trigger Backdoor Attacks (HTBA) to evaluate the effects of trigger pattern~\cite{SSP20}.
We observe that the white triggers only get 7.19\% ASR on STL10, while the ASRs of trigger HTBA-10, HTBA-12, and HTBA-14 are 97.74\%, 98.05\%, 62.74\%.

Our fourth contribution is the empirical study of the defense mechanisms.
Concretely, we investigate the detection performance from three detection-level on all model supply chain phases.
Our evaluation shows that both model-level~\cite{WYSLVZZ19} and input-level~\cite{GXWCRN19} defenses can detect backdoor attacks in the downstream phase while dataset-level~\cite{TLM18} defense works well in recognizing poisoned samples in the pre-training dataset.
To our surprise, backdoor attacks in the release phase, called Type II attack in our paper, cannot be detected by all three detection-level methods, which prompts the call for more effective defenses in future research.

\section{Preliminary}

\subsection{Masked Image Modeling (MIM)}

The core idea of MIM is masking random parts of the image and then learning to reconstruct the missing parts.
It follows the autoencoder design with the transformer architecture as the building blocks to perform the task.
The input image is first cropped to patches, e.g., $16\times 16$ patches, and MIM randomly masks certain portions of patches.
The encoder then maps the unmasked patches to a latent representation and uses the decoder to predict properties of masked patches from the latent representation.
The predicted property can be the original pixels~\cite{HCXLDG21}, latent representation~\cite{WFXWYF21}, or visual tokens~\cite{BDW21, CDWXMWHLZW22}.
The objective of MIM is to minimize the difference between predicted properties and real properties of masked patches.
Generally speaking, MIM can be concluded into two categories,  tokenizer-based methods~\cite{CDWXMWHLZW22} and end-to-end methods~\cite{ZZSYZK22}.

\mypara{Tokenizer-Based MIM}
Inspired by the success of masked language modeling, tokenizer-based MIM mimics BERT~\cite{DCLT19} to reconstruct visual tokens.
It includes two steps: utilizes an image tokenizer to generate tokens of masked patches and then optimizes the loss by predicting the correct tokens via visual patches.

\mypara{End-to-End MIM}
As the name implies, end-to-end MIM is a one-stage method without the pre-trained tokenizer.
The method is straightforward and effective.
By directly predicting large portions of masked patches with the help of small portions of unmasked patches, it can achieve impressive performance.

\subsection{Supply Chain of Self-Supervised Models}

\begin{figure}[!t]
\centering
\includegraphics[width=\linewidth]{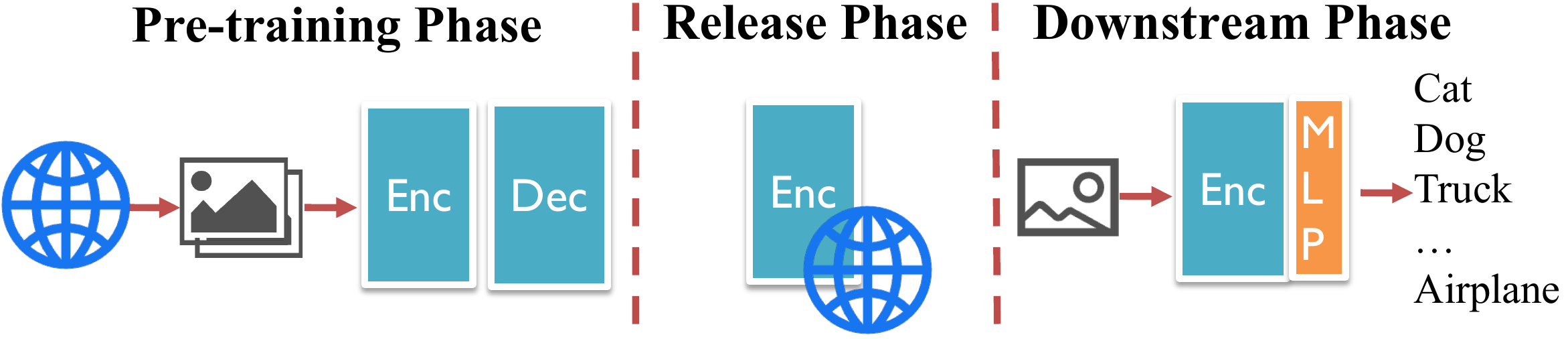}
\caption{Supply chain of self-supervised models.}
\label{figure:supply_chain}
\end{figure}

As~\autoref{figure:supply_chain} displays, the supply chain of self-supervised models can be generally summarized into three phases.
The first phase is the pre-training phase, where the model owner utilizes images collected by the data donor to train the self-supervised model.
The second phase is the release phase where the model owner makes the trained model available online via public platforms such as ModelZoo~\footnote{\url{https://modelzoo.co/}} and Hugging Face~\footnote{\url{https://huggingface.co/}}.
The third phase is the downstream phase.
In this phase, the downstream model owner adopts the pre-trained encoder as the backbone and fine-tunes an extra classification layer, i.e., MLP layer, to perform the downstream tasks.
The new model (containing an encoder and a classifier) is called \textit{downstream model}.

\subsection{Backdoor Attacks}
\label{section:backdoor_attacks}

In general, backdoor attacks inject hidden backdoors into machine learning models so that the infected models perform well on clean images but misclassify images with a specific trigger into a target class.
As an emerging and rapidly growing research area, various backdoor attacks have been proposed~\cite{GDG17, CT21, STKP21, SSP20, WSRVASLP20, LMBL20, JLG22} and can be broadly summarized into two categories, i.e., poisoning-based and non-poisoning-based backdoor attacks~\cite{LWJLX20}.

\mypara{Poisoning-based Backdoor Attack}
Give a training set $(X, Y) \in \mathcal{D}_{train}$, we first denote a target model as $f: X \rightarrow Y$ where $X \subset \mathbb{R} ^{d}$ is a set of data samples and $Y=\{1,2,...,K\}$ is a set of labels.
Given a sample $x$ with its label $y$, we assume the adversary has a target label $\widetilde{y}$ and a trigger patch $t$.
The attacker constructs a poisoned pair $(\widetilde{x}, \widetilde{y})$  by replacing the label $y$ to $\widetilde{y}$ and pasting the trigger $t$ on the image $x$ to get the patched image $\widetilde{x}$.
Then, the attacker injects a portion $p$ of poisoned pair $(\widetilde{x}, \widetilde{y})$ into $\mathcal{D}_{train}$ ( $0 < p < 1$).
Since the victim is not aware that the training set has been modified, the backdoor would be successfully embedded in the model after the training process.

\mypara{Non-poisoning-based Backdoor Attacks}
Different from poisoning-based backdoor attacks, non-poisoning-based backdoor attacks~\cite{RHF20, JLG22} directly modify model parameters to inject backdoors without poisoning the training set.
Given a clean model $f$, the attacker aims to optimize it to a backdoored model $f'$.
Concretely, the attacker collects a shadow dataset $\mathcal{D}_{shadow}$ poisoned with trigger $t$ and adopt a reference image $r$ from the target class $\widetilde{y}$.
The optimization problem aims to minimize the distance between $\mathcal{D}_{shadow}$ and $r$.

\section{Attack Taxonomy and Methodology}
\label{section:attack_taxonomy}

\begin{table*}[!t]
\centering
\caption{Attack Taxonomy. The attacks are increasingly harder in row order.  \fullcirc: Applicable or Necessary; \emptycirc: Inapplicable or Unnecessary; \halfcirc: Partially Applicable or Necessary.}
\label{table:attack_taxonomy}
\scalebox{0.8}{
\begin{tabular}{c|c|c|ccc}
\toprule
Phase $\rightarrow$     & \multicolumn{1}{c|}{Pre-training}        & Release     & \multicolumn{3}{|c}{Downstream}                                          \\
Attack $\downarrow$, Capability $\rightarrow$ & Pre-training set   & Model       & Downstream set       & Downstream model          & Inference pipeline    \\ 
\midrule
Type I                                        & \emptycirc         & \emptycirc  & \halfcirc            & \emptycirc                & \emptycirc            \\
Type II                                       & \emptycirc         & \fullcirc   & \emptycirc           & \emptycirc                & \emptycirc            \\
Type III                                      & \halfcirc          & \emptycirc  & \emptycirc           & \emptycirc                & \emptycirc            \\
\bottomrule
\end{tabular}
}
\end{table*}

As we are the first to investigate backdoor attacks on masked image modeling, we begin by defining our adversary's goal with a unified attack taxonomy covering all phases in the model supply chain.
Note, the attack taxonomy can also be generally extended to self-supervised models.

\mypara{Adversary's Goal}
Following previous work~\cite{GDG17, JLG22}, we assume the adversary aims to backdoor the downstream model so that the model performs well on clean images but misclassifies images with a specific trigger into a target class.
To achieve this goal, the adversary can perform backdoor attacks from different phases in MIM model's supply chain.

\mypara{Attack Taxonomy and Adversary's Capability}
Different from previous work, we are the first to systematically threat modeling on MIM in every phase of model supply chain, i.e., pre-training, release, and downstream phases.
\autoref{table:attack_taxonomy} shows our proposed attack taxonomy and the attacker's corresponding capabilities.
We name the backdoor attacks in each phase as Type I, Type II, and Type III attacks, respectively, and adopt three representative backdoor attacks~\cite{GDG17, JLG22, STKP21} as well as our proposed method to quantify the security risk of each phase.

\emph{Type I attack} is a poisoning-based backdoor attack that happens at the downstream phase.
We assume that the adversary knows the downstream tasks and has capability to inject a small number of labeled poisoned samples into the downstream training set.
However, they have no knowledge of pre-trained model and pre-trained dataset.
Concretely, given a downstream training set $(X, Y) \in \mathcal{D}_{down}$ and downstream classifier $\mathcal{F}$, Type I attack poisons $p$ portion of samples with trigger $t$ in $\mathcal{D}_{down}$.
The victim then uses the poisoned downstream dataset $\widetilde{\mathcal{D}}_{down}$ to optimize the downstream model.

\emph{Type II attack} is a non-poisoning-based backdoor attack and takes place in the release phase.
The attacker can be either an untrusted service provider who injects a backdoor into its pre-trained model or a malicious third-party who downloads the released pre-trained model, injects a backdoor into it, and then re-publishes it online~\cite{JLG22}.
In this scenario, the attacker has full access to the pre-trained model but has no knowledge of the pre-training dataset, downstream dataset, and downstream training schedule.
Specifically, given a clean MIM model $\mathcal{M}$, we have $\hat{x} = \mathcal{M}(x) = Dec(Enc(x))$, where $Enc$ is the encoder and $Dec$ is the decoder.
To train a downstream task, the decoder $Dec$ will be discard and the victim will build a new model $\mathcal{F}$ so that $\hat{y} = \mathcal{F}(x) = MLP(Enc(x))$.
The goal of attacker is to optimize $Enc$ to a poisoned $\widetilde{Enc}$ so that $\widetilde{y} = \widetilde{\mathcal{F}}(x) = MLP(\widetilde{Enc}(\widetilde{x}))$ where $\widetilde{y}$ is the target class and $\widetilde{x}$ is a poisoned sample.

\emph{Type III attack} is a poisoning-based backdoor attack.
Similar to Type I attack, the attackers have no knowledge of the model hyperparameters and can only poison a small fraction of the pre-training dataset.
However, unlike Type I attack where the attacker can directly change the label of poisoned samples in the downstream dataset, the pre-training dataset has no label.
To address this issue, the attacker in Type III attack only poisons samples from the target class by adding triggers to them and expects the pre-trained model to recognize the triggers as a part of the target class to establish an inner connection between the trigger and the specific target class.
In reality, Type III attacker can be a malicious data donor who releases poisoned images on the Internet.
Once the poisoned images are scraped by the model owner without censoring, they can inject backdoors into the pre-trained models.

\section{Evaluation}

\subsection{Experimental Settings}
\label{section:exp_details}

\mypara{Datasets}
We utilize four datasets in our experiments.
For Type I and Type II attacks, we use publicly available ImageNet pre-trained MIM models and use CIFAR10, CIFAR100, and STL10 as the datasets to perform the downstream tasks.
For Type III attack, we use ImageNet20 to pre-train MIM models and consider CIFAR10, STL10, and ImageNet20 as the downstream datasets.
All images are resized to $224\times 224$ to fit the input requirement of the models, which is also a common practice in related work~\cite{JLG22, DBKWZUDMHGUH21}.

\mypara{Target Model}
We consider two MIM architectures as the target models, i.e., Masked Autoencoder (MAE)~\cite{HCXLDG21} for end-to-end MIM and Contextual Autoencoder (CAE)~\cite{CDWXMWHLZW22} for tokenizer-based MIM.
For both the two target models, we adopt the same base variant of ViT (ViT-B) with 224 × 224 input image size and 16 × 16 patch size.

Concretely, for Type I and Type II attacks, as the adversary does not involve in the pre-training phase, we utilize the public MAE~\footnote{\url{https://github.com/facebookresearch/mae}} and CAE~\footnote{\url{https://github.com/lxtGH/CAE}} as our target model.
This aligns with the threat model that attackers can only get access to the released models.
For Type III attack, we use ImageNet dataset to train the two target models from scratch.
Note, the models contain around 89M and 149M parameters, which costs huge time and computing resources to train it on the complete ImageNet dataset from scratch.
Therefore, we instead use a subset of ImageNet to perform a quick evaluation in the pre-training phase.
The subset contains 20 randomly-extract labels (see~\autoref{table:imagenet20} in Appendix).
This is also a common way to do the evaluation~\cite{STKP21,TKI202}.
Note that in Type III attack, we replace the CIFAR100 with ImageNet20 as the downstream dataset as the pre-training dataset ImageNet20 does not cover all classes on CIFAR100, which yields less satisfying clean accuracy.
Also, previous work~\cite{HCXLDG21, CDWXMWHLZW22} leverages the pre-training dataset as the downstream dataset as well.

\mypara{Metric}
We consider four evaluation metrics.
Test accuracy (TA)/clean accuracy (CA) measures the classification accuracy of the backdoored/clean model on clean testing images.
Attack success rate (ASR)/attack success rate-baseline (ASR-B) denotes the classification accuracy of the backdoored/clean model on poisoned testing images with triggers.

We refer the readers to~\autoref{section:exp_details_appendix} for detailed descriptions of the datasets, triggers, and configurations of pre-training tasks, downstream tasks, backdoor attacks, and defense methods.

\subsection{Main Experiment Results}

\begin{table*}[!t]
\centering
\caption{Attack performance on MAE and CAE (shown with percentage).
Type III-R is adapted from Saha et al.\ where the trigger is randomly placed on the images~\cite{STKP21}.
Type III-M is the method proposed in this paper where we put nine same triggers on the images against the impacts of masking.
The clean accuracy (CA) in Type III attack is lower than the other two scenarios, that is because it is trained on the subset of ImageNet.}
\label{table:overall_backdoor}
\scalebox{0.8}{
\begin{tabular}{lll|cccc|cccc|cccc}
\toprule
\multirow{2}{*}{\textbf{Phase}} & \multirow{2}{*}{\textbf{Attack}} & \multirow{2}{*}{\textbf{Model}} & \multicolumn{4}{|c}{\textbf{CIFAR10}} & \multicolumn{4}{|c}{\textbf{CIFAR100}} & \multicolumn{4}{|c}{\textbf{STL10}} \\
             &            &     & TA       & CA        & ASR   & ASR-B & TA       & CA        & ASR   & ASR-B & TA       & CA        & ASR   & ASR-B \\ 
\midrule
Downstream   & Type I     & MAE & 86.64    & 87.73     & 99.62 & 10.00 & 63.69    & 68.30     & 98.74 & 1.00  & 92.63    & 95.05     & 97.40 & 10.00 \\
Release      & Type II    & MAE & 87.62    & 85.49     & 96.48 & 10.00 & 67.86    & 68.30     & 67.57 & 1.00  & 94.61    & 95.05     & 99.18 & 10.00 \\
Pre-training & Type III-R & MAE & 69.36    & 68.95     & 53.04 & 50.21 & 42.54    & 42.30     & 19.75 & 4.34  & 62.73    & 62.83     & 28.88 & 15.10 \\
Pre-training & Type III-M & MAE & 69.32    & 68.98     & 98.89 & 57.04 & 42.52    & 42.30     & 19.44 & 1.84  & 62.39    & 65.58     & 97.74 & 17.51 \\ 
\midrule
Downstream   & Type I     & CAE & 92.25    & 93.41     & 99.58 & 11.68 & 73.09    & 77.46     & 98.64 & 0.69  & 93.63    & 96.31     & 97.99 & 9.95  \\
Release      & Type II    & CAE & 90.05    & 93.41     & 90.01 & 11.68 & 71.86    & 77.46     & 66.34 & 0.69  & 93.75    & 96.31     & 95.88 & 9.95  \\
Pre-training & Type III-R & CAE & 70.74    & 67.80     & 26.58 & 28.29 & 45.96    & 42.70     & 8.61  & 6.05  & 62.80    & 62.20     & 12.46 & 11.68 \\
Pre-training & Type III-M & CAE & 70.49    & 67.80     & 51.95 & 34.54 & 45.22    & 42.70     & 14.14 & 8.97  & 64.16    & 62.20     & 15.66 & 12.53 \\
\bottomrule
\end{tabular}
}
\end{table*}

\autoref{table:overall_backdoor} shows the performance of backdoor attacks in all three phases on models built with MIM.

\mypara{Type I Attack (Downstream Phase)}
Overall, we observe that the downstream phase is the most fragile phase in the supply chain of MIM.
For all downstream tasks and target models, the backdoor attack can reach extremely high ASR.
For instance, the ASR of Type I attack are 99.62\%, 98.74\%, and 97.40\% on CIFAR10, CIFAR100, and STL10, respectively.

\mypara{Type II Attack (Release Phase)}
To compare, if the attack occurs in the release phase, the effect of the attack is relatively unstable because the attacker has no knowledge of the downstream phase.
However, this phase is still vulnerable to backdoor attacks.
We observe the ASR ranged from 66.34\% to 99.18\% on both MAE and CAE.

\mypara{Type III Attack (Pre-training Phase)}
From the perspective of the attacker, Type III attack is the hardest attacking scenario.
In this scenario, the model is trained on an unlabeled dataset.
Therefore, the attacker cannot directly associate the trigger with the target label.
Attacks in this scenario have never been thoroughly explored in previous studies.
To the best of our knowledge, only Saha et al.\ investigate backdoor attack in this scenario, which is Type III-R attack~\cite{STKP21}.
It randomly puts a single trigger on the images of the target class.
However, we observe it can not achieve satisfying attack performance on models built with MIM.
The ASR is only 2.83\% and 13.78\% higher than the baseline on CIFAR10 and STL10, respectively.
The reason behind this could be credited to the masking mechanism.
As MIM methods randomly mask a large portion of the input images, i.e., 75\% in MAE, the trigger can be masked in the pre-training phase.
Intuitively, we propose Type III-M attack to improve the attack performance in this scenario, in which nine same triggers are put on the images to alleviate the impacts of masking.
We observe that by increasing the number of triggers, the ASR can mount to 98.89\% and 97.74\% on CIFAR10 and STL10 datasets in the end, which outperforms Type III-R attack significantly.
Besides, we find that backdoor attacks occurring in the per-training phase can preserve the utility of the model to a large extent.
Take CAE as an example.
The test accuracy on CIFAR10 is 70.74\% and 70.49\% for Type III-R attack and Type III-M attack, which are even 2.94\% and 2.69\% higher than the clean accuracy, respectively.

\section{What Make Each Phase Different}
\label{section:ablation_study_main}

We then take MAE as the target model's architecture and conduct comprehensive ablation studies to understand the impacts of important backdoor attack components in each supply chain's phase.
We report our main and intriguing findings here and refer the readers to~\autoref{section:ablation_study_appendix} for detailed experiment results.

\begin{figure*}[!t]
\centering
\includegraphics[width=0.8\linewidth]{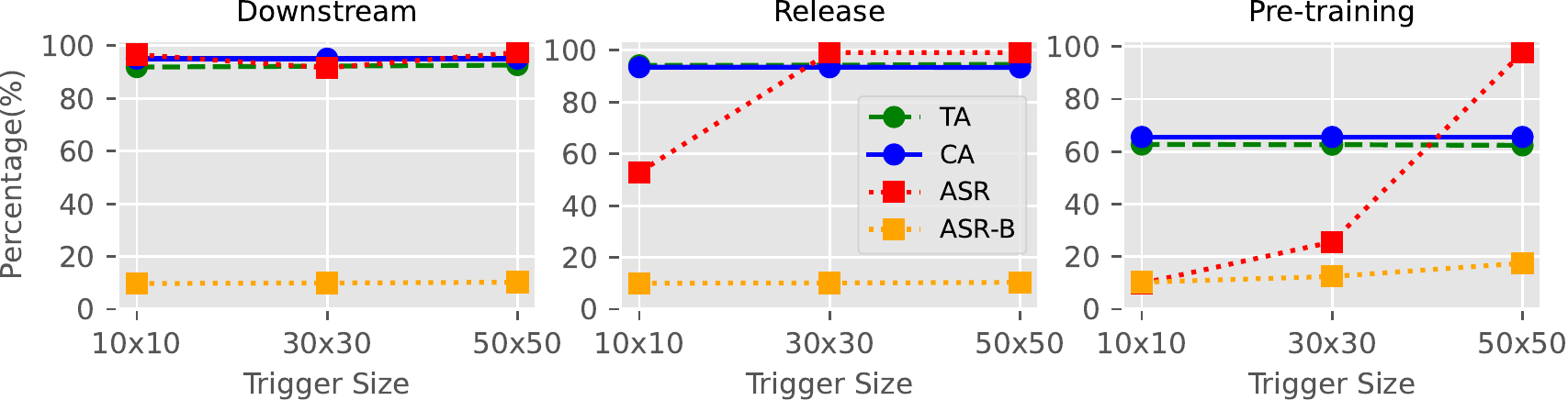}
\caption{Impact of trigger size in different model supply chain phases on STL10.}
\label{figure:trigger_size_stl10}
\end{figure*}

\mypara{Impacts of Trigger Size at different phases}
\autoref{figure:trigger_size_stl10} and~\autoref{figure:trigger_size_cifar10} (in \autoref{section:ablation_study_appendix}) show the performance under different trigger size.
Interestingly, we observe a clear but distinguishable increase when trigger size enlarges in different phases, which indicates that a larger trigger can achieve better performance and backdooring pre-training phase is harder than release and downstream phases.

\begin{figure*}[!t]
\centering
\includegraphics[width=0.8\linewidth]{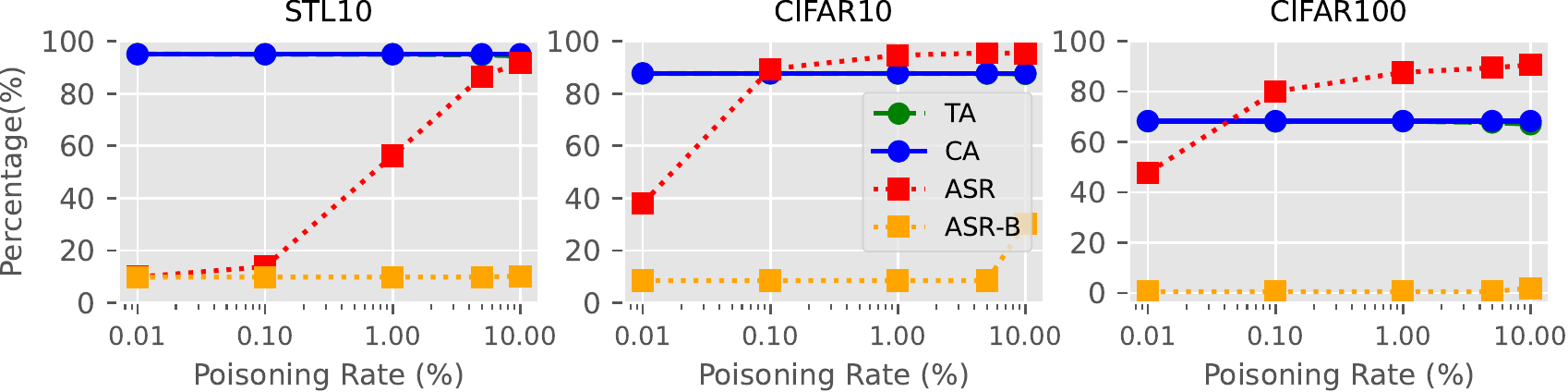}
\caption{Impacts of poisoning rate in Type I attack. Trigger size is $5\times 5$.}
\label{figure:att_1_poison_rate_tri5}
\end{figure*}

\mypara{The Fragility of the Downstream Phase}
Based on the results of trigger size (\autoref{figure:att_1_trigger_size}) and poisoning rate (\autoref{figure:att_1_poison_rate_tri50}) in \autoref{section:ablation_study_appendix}, we observe the downstream phase is extremely vulnerable to backdoor attacks.
For instance, the attacker can achieve 93.80\% ASR when the poisoning rate is only 10\%.
To test the limits of this attack, we continue to reduce trigger size from $50\times 50$ to $5\times 5$ and decrease the poisoning rate, as~\autoref{figure:att_1_poison_rate_tri5} shows.
Take CIFAR10 as an example. 
With only 0.1\% poisoning rate (e.g., 50 training samples) and 0.05\% occupied area of the image (e.g., $5\times 5$), the attacker still achieves 89.37\% ASR.
And when the poisoning rate is extremely low, i.e., 0.01\% poisoning rate (e.g., only 5 training samples), the attacker can still conduct backdoor attacks successfully, indicating the fragility of the downstream phase.
We attribute this vulnerability to the powerful representative capability of MIM and also the capability of the attacker, i.e., they can directly get access to the downstream dataset.

\begin{table}[!t]
\centering
\caption{Impacts of mask. ``Without Mask'' means the encoder output latent of all patches. ``With Mask'' means the encoder would randomly mask 75\% patches and only output latent of visible patches.}
\label{table:att_2_mask}
\scalebox{0.8}{
\begin{tabular}{l|cc|cc|cc}
\toprule
         & \multicolumn{2}{c|}{\textbf{Without Mask}} & \multicolumn{2}{c|}{\textbf{With Mask}} & \multicolumn{2}{c}{\textbf{Clean Model}} \\ 
         & TA               & ASR           & TA              & ASR         & CA               & ASR-B        \\ 
\midrule
STL10    & 94.61            & 99.18         & 47.00           & 0.00        & 93.40            & 10.00        \\
CIFAR10  & 87.62            & 96.48         & 46.88           & 27.19       & 85.49            & 10.00        \\
CIFAR100 & 67.86            & 67.57         & 20.11           & 0.11        & 63.55            & 1.00         \\
\bottomrule       
\end{tabular}
}
\end{table}

\mypara{Mask Mechanism Is a Stumbling Block to Type II Attack}
Mask is a key component of MAE.
By randomly masking a portion of patches and optimizing the loss between reconstructed masked patches and real patches, MAE achieves state-of-the-art performance.
Conventionally, after obtaining the released MAE model, Type II attacker would directly apply backdoor attacks on the encoder.
However, our experiments show that only by removing the mask component while attacking, the backdoor can be successfully embedded (the removed mask component can be added back after the attack is finished).
\autoref{table:att_2_mask} shows the attack performance of Type II attack without mask and with mask.
It is clear that backdoor attack cannot work well with mask mechanism.
We believe that the results are due to the fact that the masking mechanism causes the patches from the backdoor model and the clean model to be misaligned.
In detail, as Type II attack needs to calculate the loss of patches between clean model and backdoored model, the randomness of masking will distort the feature space of the model.

\mypara{The Success Factors of Type III Attack}
To the best of our knowledge, pre-training phase, as the hardest scenario, has never been thoroughly explored in previous studies.
To fill this gap, we conduct comprehensive ablation studies on the poisoning rate as well as the properties of triggers, i.e., pattern, location, number, and size.

We find that trigger pattern and trigger number are key factors that affect attack performance in the pre-training phase while trigger location has limited impact, which is different from a previous study on contrastive learning~\cite{STKP21}.
\autoref{table:att_3_trigger_pattern} shows the experimental results and~\autoref{figure:trigger} displays the triggers.
We observe that the white triggers only get 7.19\% ASR on STL10, while the ASRs of trigger HTBA-10, HTBA-12, and HTBA-14 are 97.74\%, 98.05\%, 62.74\%, respectively.
One possible reason is that self-supervised models have no label.
Therefore, it's hard for the model to directly connect the trigger to target classes.
We remain the reason behind vary attack performance of different trigger patterns for future work.

We then test four different trigger putting methods to poison the pre-training dataset, i.e., random, localization, center, and multiple.
The results are shown in~\autoref{table:trigger_location}.
Surprisingly, we find that the success of Type III attack is mainly related to trigger number rather than trigger location or whether the trigger appears on the target object.
For example, the ASR of random, localization, and center methods are 28.88\%, 26.88\%, and 26.66\% on STL10, respectively.
However, when trigger occurrence number increase, the ASR increases to 97.74\%.

\begin{figure*}[!t]
\centering
\includegraphics[width=0.8\linewidth]{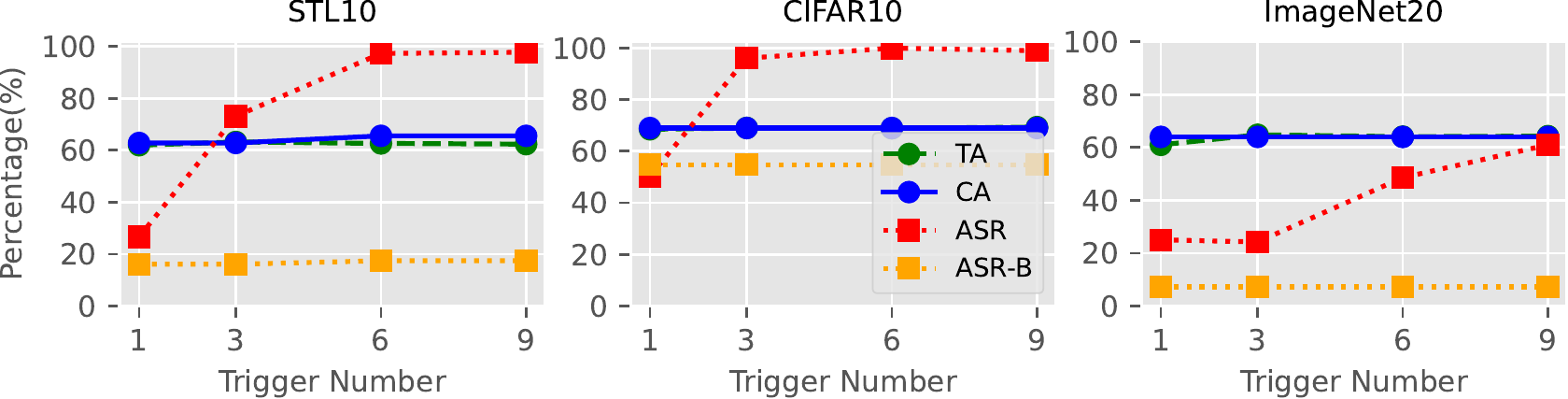}
\caption{Impacts of trigger number in Type III-M attack.}
\label{figure:att_3_tri_num}
\end{figure*}

With the following experiments on trigger numbers (see~\autoref{figure:att_3_tri_num}), we found that by increasing the number of trigger, we can effectively bypass the masking process.
For example, when trigger number is 3, we can already achieve 95.97\% ASR on CIFAR10.

\begin{table}[!t]
\centering
\caption{Impacts of trigger pattern. Triggers can be found in~\autoref{figure:trigger}.}
\label{table:att_3_trigger_pattern}
\scalebox{0.8}{
\begin{tabular}{c|l|cc|cc}
\toprule
\textbf{Trigger Pattern} & \textbf{Dataset} & \textbf{TA}       & \textbf{CA}       & \textbf{ASR}    & \textbf{ASR-B}  \\ 
\midrule
\multirow{3}{*}{White}   & STL10            & 61.98             & 62.83              & 7.19           & 13.36           \\
                         & CIFAR10          & 68.58             & 68.95              & 26.49          & 34.75           \\
                         & ImageNet         & 61.70             & 63.20              & 1.90           &  6.30           \\ 
\midrule
\multirow{3}{*}{HTBA-10} & STL10            & 62.39             & 65.58              & 97.74          & 17.51           \\
                         & CIFAR10          & 69.32             & 68.98              & \textbf{98.89} & 57.04           \\
                         & ImageNet         & 64.30             & 63.20              & 61.00          &  8.10           \\ 
\midrule
\multirow{3}{*}{HTBA-12} & STL10            & 63.09             & 65.58              & \textbf{98.05} & 16.30           \\
                         & CIFAR10          & 69.55             & 68.98              & 61.96          & 53.78           \\
                         & ImageNet         & 62.50             & 63.20              & \textbf{70.80} &  6.60           \\ 
\midrule
\multirow{3}{*}{HTBA-14} & STL10            & 63.00             & 65.58              & 62.74          & 19.10           \\
                         & CIFAR10          & 69.36             & 68.98              & 0.00           & 57.61           \\
                         & ImageNet         & 63.60             & 63.20              & 0.00           &  8.00           \\
\bottomrule
\end{tabular}
}
\end{table}

\begin{table}[!t]
\caption{ Impacts of trigger location. \autoref{figure:trigger_location} displays examples of trigger put methods.}
\label{table:trigger_location}
\centering
\scalebox{0.8}{
\begin{tabular}{c|l|cc|cc}
\toprule
\multicolumn{1}{l|}{\textbf{Trigger Position}}  & \textbf{Dataset}  & \textbf{TA} & \textbf{CA} & \textbf{ASR}   & \textbf{ASR-B}   \\ 
\midrule
\multirow{3}{*}{Random}                         & STL10             & 62.73       & 62.83       & 28.88          & 15.10            \\
                                                & CIFAR10           & 69.36       & 68.95       & 53.04          & 50.21            \\
                                                & ImageNet          & 63.90       & 64.00       & 21.40          &  9.90            \\ 
\midrule
\multirow{3}{*}{Localization}                   & STL10             & 62.78       & 62.83       & 26.88          & 15.10            \\
                                                & CIFAR10           & 69.18       & 68.95       & 53.85          & 50.21            \\
                                                & ImageNet          & 63.80       & 64.00       & 20.40          &  9.90            \\ 
\midrule
\multirow{3}{*}{Center}                         & STL10             & 61.98       & 65.58       & 26.66          & 17.51            \\
                                                & CIFAR10           & 68.46       & 68.98       & 50.33          & 57.04            \\
                                                & ImageNet          & 61.00       & 63.20       & 25.10          &  8.10            \\ 
\midrule
\multirow{3}{*}{Multiple}                       & STL10             & 62.39       & 65.58       & \textbf{97.74} & 17.51            \\
                                                & CIFAR10           & 69.32       & 68.98       & \textbf{98.89} & 57.04            \\
                                                & ImageNet          & 64.30       & 63.20       & \textbf{61.00} &  8.10            \\
\bottomrule
\end{tabular}
}
\end{table}

\section{Can Current Defense Mitigate Backdoor Attacks}

Many methods have been proposed to defend against backdoor attacks~\cite{XWLBGL21, WYSLVZZ19, GXWCRN19, TLM18}.
Overall, they can be categorized into three detection levels~\cite{XWLBGL21}, i.e., model-level, input-level, and dataset-level.
We evaluate the performance of backdoor attacks under all scenarios in all detection levels.
For each detection level, we select one of the most representative methods.
Our evaluation shows that both model-level~\cite{WYSLVZZ19} and input-level~\cite{GXWCRN19} defenses can detect backdoor attacks in the downstream phase while dataset-level~\cite{TLM18} defense works well in recognizing poisoned samples in the pre-training dataset.

To our surprise, backdoor attacks in the release phase, called Type II attack in our paper, cannot be detected by all three detection-level methods, which calls for future research.

\mypara{Model-level Defense}
Given a classifier, Neural Cleanse~\cite{WYSLVZZ19} calculates the anomaly index to identify whether it is backdoored or not.
We follow the default parameter settings of Neural Cleanse and conduct it on the downstream models of all three attacks.
\autoref{table:neural_cleanse} shows the anomaly indices and predicted target label of Neural Cleanse.
A model is predicted to be backdoored if anomaly index is higher than 2.
If predicted target label is correct,``Pred'' is filled by $\checkmark$.
We observe that Neural Cleanse performs well on Type I attack.
The anomaly index for CIFAR10 and STL10 are 2.27 and 2.15, respectively.
The predicted target label is also correct.
However, for Type II and Type III-M attacks, the anomaly scores are lower than 2, indicating that Neural Cleanse cannot detect backdoors embedded in release and pre-training phases.

\begin{table}[!t]
\caption{Anomaly Indices produced by Neural Cleanse.}
\label{table:neural_cleanse}
\centering
\scalebox{0.8}{
\begin{tabular}{l|cc|cc}
\toprule
           & \multicolumn{2}{c|}{\textbf{CIFAR10}} & \multicolumn{2}{c}{\textbf{STL10}} \\
& \multicolumn{1}{|l}{Index}    & \multicolumn{1}{l|}{Pred}     & \multicolumn{1}{|l}{Index}    & \multicolumn{1}{l}{Pred}  \\ 
\midrule
Type I      & 2.27              & $\checkmark$                  & 2.15                          & $\checkmark$              \\
Type II     & 0.82              & -                             & 1.48                          & -                         \\
Type III-M  & 1.57              & -                             & 1.97                          & -                         \\
\bottomrule
\end{tabular}
}
\end{table}

\begin{figure*}[!t]
\centering
\begin{subfigure}{0.25\linewidth}
\includegraphics[width=\columnwidth]{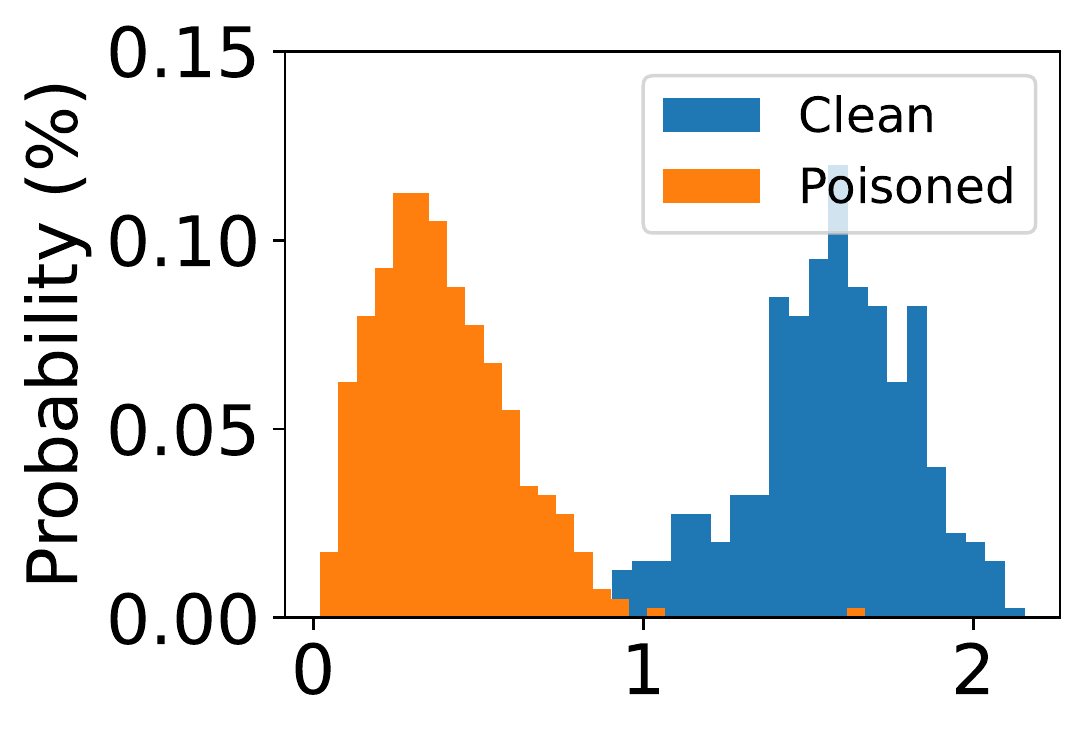}
\caption{Type I attack}
\end{subfigure}
\begin{subfigure}{0.25\linewidth}
\includegraphics[width=\columnwidth]{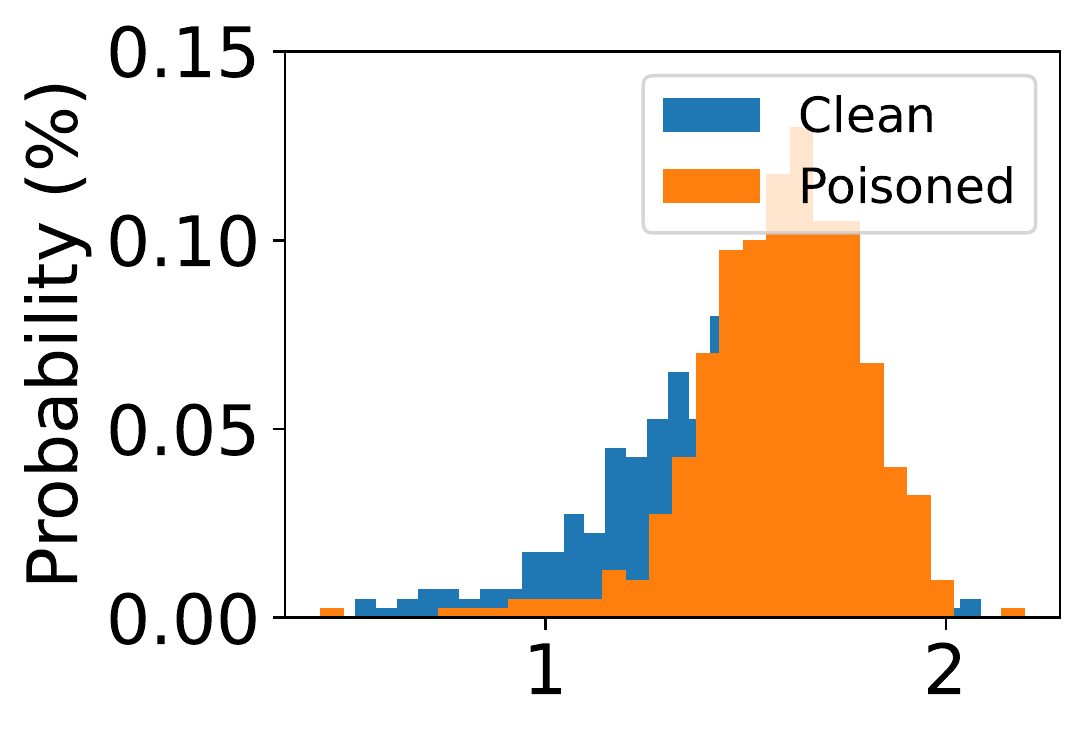}
\caption{Type II attack}
\end{subfigure}
\begin{subfigure}{0.25\linewidth}
\includegraphics[width=\columnwidth]{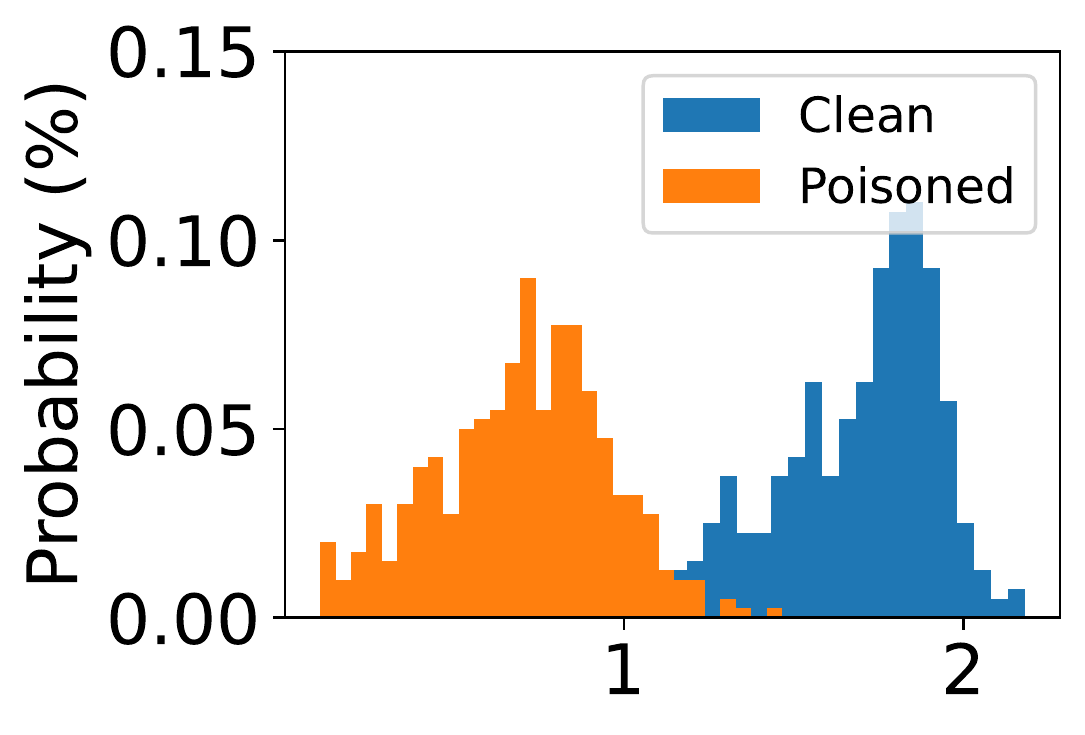}
\caption{Type III-M attack}
\end{subfigure}
\caption{Entropy distribution of CIFAR10, calculated by STRIP.}
\label{figure:strip_cifar10}
\end{figure*}

\mypara{Input-level Defense}
STRIP~\cite{GXWCRN19} is a detection method that distinguishes the testing images at run-time.
It intentionally perturbs the incoming input by blending various image patterns and calculates the entropy of the predicted classes for perturbed inputs from a given model.
A low entropy violates the input-dependence property of a benign model and implies the presence of a perturbed input.
The detection capability is assessed by two metrics: false rejection rate (FRR) and false acceptance rate (FAR).
The FRR is the probability when the benign input is regarded as a poisoned input.
The FAR is the probability that the poisoned input is recognized as the benign input.
Ideally, both FRR and FAR should be 0\%.

\autoref{table:strip} displays the FAR and FRR of backdoored models in the three attack scenarios.
We observe that the detection performance of STRIP decreased by order of Type I, Type III, and Type II attack.
For instance, for Type I attacks, the FAR and FRR are 0.75\% and 2.25\% on CIFAR10, indicating it can clearly distinguish between clean samples and poisoned samples in Type I attack.
To compare, the FAR and FRR for Type II (Type III-M) attack are 99.50\% and 3.00\% (11.00\% and 4.25\%).
To further understand the failure reason of STRIP, we visualize the entropy distribution in~\autoref{figure:strip_cifar10} and~\autoref{figure:strip_stl10} (in Appendix).
We observe that STRIP fails to distinguish between poisoned images and clean images in Type II attack.
One possible reason is that Type II attack tends to drag the feature space of poisoned samples to the reference images.
Therefore, STRIP is likely to regard the perturbed samples as reference images, which is still clean sample.
Besides, we observe that STRIP can distinguish part of poisoned samples from Type III-M attack.
However, the performance is not stable, i.e., it can distinguish perturbed samples from CIFAR10 but fail in STL10 (see ~\autoref{figure:strip_stl10}).
This instability has also been shown in other works~\cite{SWBMZ22, GCDW22}.

\begin{table}[!t]
\centering
\caption{FRR and FAR of STRIP. If both FRR and FAR are 0, STRIP can be regarded as a good detection.}
\label{table:strip}
\scalebox{0.8}{
\begin{tabular}{l|cc|cc}
\toprule
 & \multicolumn{2}{c|}{\textbf{CIFAR10}} & \multicolumn{2}{c}{\textbf{STL10}} \\
 & \multicolumn{1}{c}{FAR (\%)} & \multicolumn{1}{c|}{FRR (\%)} & \multicolumn{1}{c}{FAR (\%)}  & \multicolumn{1}{c}{FRR (\%)}  \\ 
\midrule
Type I      & 0.75              & 2.25                          & 7.50                          & 2.19                          \\
Type II     & 99.50             & 3.00                          & 99.38                         & 0.94                          \\
Type III-M  & 11.00             & 4.25                          & 94.06                         & 1.56                          \\
\bottomrule
\end{tabular}
}
\end{table}

\mypara{Dataset-level Defense}
Spectral signatures~\cite{TLM18} defend poisoning-based backdoor attacks at the dataset level.
It assumes attackers tend to poison a subset of training set to inject backdoors in the model, which might lead to detectable traces in the covariance spectrum of the poisoned and clean feature representation.
By calculating the outlier score of the feature representation, spectral signatures can detect and remove poisoned images from the training set.
However, for backdoor attacks that involve no data poisoning, i.e., Type II attack, spectral signature is not a suitable defense method.
\autoref{table:spectral_sign} shows both the backdoor score and clean score from spectral signatures.
We observe that it can clearly detect poisoned samples in Type I and Type III attacks.
However, as~\autoref{figure:att_1_poison_rate_tri5} shows, Type I attack can still achieve high ASR when the poisoning rate is quite low.
And under this situation, we find that spectral signatures start losing efficacy.
For instance, when the poisoning rate is 1\%, which is 500 images in CIFAR10 dataset, the backdoor score is lower than the clean score, showing spectral signature cannot distinguish the poisoned images.

\begin{table}[!t]
\centering
\caption{Scores of spectral signature.
B-Score/C-Score refers to backdoor/clean score.
If B-Score $>$ C-Score, spectral signature can identify poisoned samples.}
\label{table:spectral_sign}
\scalebox{0.8}{
\begin{tabular}{l|l|c|c|cc}
\toprule
\textbf{Attack}         & \textbf{Trainset} & \multicolumn{1}{l|}{\textbf{Poi (\%)}} & \multicolumn{1}{l|}{\textbf{ASR}} & \multicolumn{1}{l}{\textbf{B-Score}} & \multicolumn{1}{l}{\textbf{C-Score}} \\ 
\midrule
\multirow{2}{*}{Type I} & CIFAR10           & 50.00                      & 99.62                     & 7.83          & 5.67      \\
                        & STL10             & 50.00                      & 97.40                     & 10.87         & 7.79      \\ 
\midrule
\multirow{2}{*}{Type I} & CIFAR10           & 1.00                       & 94.67                     & 5.55          & 7.82      \\
                        & STL10             & 1.00                       & 56.23                     & 3.49          & 6.71      \\ 
\midrule
Type II                 & -                 & \multicolumn{1}{l|}{-}     & \multicolumn{1}{l|}{-}    & -             & -         \\ 
\midrule
Type III-M              & ImageNet          & 4.50                       & 98.89                     & 7.51          & 4.31      \\
\bottomrule
\end{tabular}
}
\end{table}

\section{Limitations}

In this paper, we focus on backdoor attacks against MIM among the whole supply chain and apply the most representative backdoor attack methods in each phase.
However, there are also some advanced backdoor attacks that use dynamic trigger~\cite{SWBMZ22}, hidden trigger~\cite{SSP20}, or attack multiple target labels simultaneously.
We leave them as our future work for further exploration.

\section{Related Work}

\mypara{Backdoor Attacks Against Pre-trained Models}
Various machine learning models are shown to be vulnerable to backdoor attacks, i.e., deep neural networks~\cite{GDG17}, graph neural networks~\cite{XPJW21}, and federated learning~\cite{XHCL20}.
Among them, Jia et al.\ first proposed backdoor attacks against pre-trained encoders~\cite{JLG22}.
Then, Carlini and Terzis~\cite{CT21} proves backdoor attacks on supervised learning can be directly adopted on pre-trained models.
Saha et al.\ challenged the hardest setting, whereby the attacker can only poison the self-supervised training set~\cite{STKP21}.
However, all of the above backdoor attacks against pre-trained models are mainly focused on contrastive learning-based models, a discriminative method.
In contrast, masked image modeling, as a generative method showing remarkable performance recently, has never been systematically studied.
Thus, we take the first step to quantify backdoor attacks on models built by masked image modeling.

\mypara{Defense of Backdoor Attacks}
Many methods have been proposed to defend against backdoor attacks~\cite{XWLBGL21, WYSLVZZ19, GXWCRN19, TLM18}.
Overall, they can be categorized into three detection levels~\cite{XWLBGL21}, i.e., model-level, input-level, and dataset-level.
Wang et al.\ proposed the first defense method against backdoor attacks on deep neural networks~\cite{WYSLVZZ19}.
By finding the label that requires smaller modifications to cause misclassification on a specific target class, it achieves model-level backdoor detection.
Instead of identifying injected models, STRIP~\cite{GXWCRN19} filters out inputs in the inference time to brake backdoor activation by distinguishing entropy distribution of perturbed samples and clean samples.
The dataset-level detection settles at the beginning of model training and aims to sanitize the poisoned samples from the training set.
Based on the detectable traces in the covariance spectrum of the perturbed and clean feature representation, spectral signatures~\cite{TLM18} detect poisoned images by calculating the outlier score of the feature representation.

\section{Conclusion}

In this paper, we perform the first security risk quantification of MIM through the lens of backdoor attacks.
Different from previous work, we are the first to systematically threat modeling on MIM in every phase of model supply chain, i.e., pre-training, release, and downstream phases.
Our evaluation shows that models built with MIM are vulnerable to existing backdoor attacks in release and downstream phases and are compromised by our proposed method in pre-training phase.
We also take the first step to investigate the success factors of backdoor attacks in the pre-training phase and find the trigger pattern and trigger number play key roles in the success of backdoor attacks while trigger location has tiny effects.
In the end, our empirical study of the defense mechanisms across three detection-level on model supply chain phases indicates that different defenses are suitable for backdoor attacks in different phases of MIM's supply chain.
However, backdoor attacks in the release phase cannot be detected by all three detection-level methods, calling for future research.

\bibliographystyle{plain}
\bibliography{normal_generated_py3}

\newpage
\appendix
\section{Appendix}
\label{section:appendix}

\subsection{Detailed Experimental Settings}
\label{section:exp_details_appendix}

\mypara{Datasets}
The description for datasets is as follows.
\begin{itemize}
\item \mypara{CIFAR10} 
This dataset contains 60,000 images with 10 labels.
For each label, it consists of 5,000 training images and 1,000 test images.
The size of each image is $32\times 32$ pixels.
\item \mypara{CIFAR100} 
This dataset obtains 60,000 images with 100 labels.
Each label has 600 images.
The size of each image is also $32\times 32$ pixels.
\item \mypara{STL10}
This dataset is a 10-classes image dataset.
Each class has 500 training images and 800 test images with the size of $96\times 96$.
\item \mypara{ImageNet} 
ImageNet is a common pre-training dataset in masked image modeling, containing 1,000 labels and millions of samples.
In our experiments, we adopt a 20-labels subset to do a quick evaluation on Type III attack.
The list of classes from ImageNet20 can be found in~\autoref{table:imagenet20}.
\end{itemize}

\begin{figure}[!t]
\centering
\begin{subfigure}{0.24\columnwidth}
\centering
\includegraphics[width=.6\columnwidth]{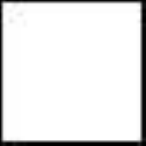}
\caption{White}
\label{figure:trigger_white}
\end{subfigure}
\begin{subfigure}{0.24\columnwidth}
\centering
\includegraphics[width=.6\columnwidth]{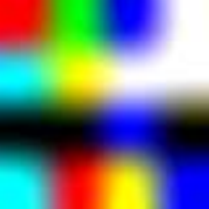}
\caption{HTBA-10}
\label{figure:trigger_10}
\end{subfigure}
\begin{subfigure}{0.24\columnwidth}
\centering
\includegraphics[width=.6\columnwidth]{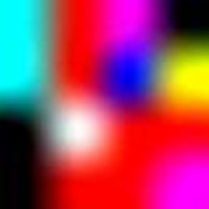}
\caption{HTBA-12}
\label{figure:trigger_12}
\end{subfigure}
\begin{subfigure}{0.24\columnwidth}
\centering
\includegraphics[width=.6\columnwidth]{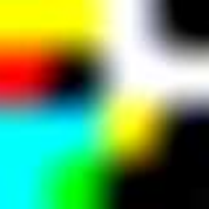}
\caption{HTBA-14}
\label{figure:trigger_14}
\end{subfigure}
\caption{Triggers. Note that (b), (c), and (d) are adopted from Saha et al.~\cite{SSP20}.}
\label{figure:trigger}
\end{figure}

\mypara{Trigger}
We use two different kinds of backdoor triggers to evaluate the performance, as~\autoref{figure:trigger} shows.
We use the white square trigger in all three scenarios, which is a common trigger used in many backdoor attacks~\cite{JLG22, GDG17}.
The other three triggers are square triggers generated by a random $4\times 4$ RGB image and then resized to desired patch size, adopted from Saha et al.\ with original ID~\cite{SSP20}.
We use these triggers in the pre-training phase to analyze the impacts of trigger patterns (see~\autoref{section:ablation_study_main}).

\mypara{Pre-training Configuration}
For MAE, the batch size is 32, epoch is 200, mask ratio is 75\%, and norm pix loss is False.
We use Adam optimizer with a base learning rate of 1.5e-4 and a warmup of 40 epochs.
The learning rate scheduler is cosine with 0.05 weight decay.
For CAE, the batch size is 32.
The base learning rate is 1.5e-3.
We use a cosine learning rate decay schedular with 0.05 weight decay.
The warmup epochs is 10 and the epoch is 100.
The drop path rate is 0.1 and dropout rate is 0.
The mask ratio is 50\%, following the default settings.

\mypara{Downstream Configuration}
To promise the results are comparable, we adopt the same linear probing configurations in all three scenarios for both MAE and CAE.
We use AdamW optimizer with weight decay 0.05, learning rate 1e-3, and a scheduler to decay it $0.9\times$ every epoch.
The model is trained for 30 epochs.
The batch size is 256.
We do not use the same optimizer of the original paper because LARS works better on large batch training and large datasets \cite{YGG17}.
However, due to the size of the downstream dataset and computing resource limits, AdamW is more suitable under a small batch size setting.
We compare the MAE performance of using AdamW, SGD, and LARS as the optimizer and find AdamW reaches the best clean accuracy (see \autoref{table:optimizer}).

\mypara{Type II Attack Configuration}
Following the experiment setting in the paper~\cite{JLG22}, we use 1\% ImageNet as the shadow model.
The trigger is put at the right bottom of the images and the size of the trigger is $50\times 50$.
We use reference images from Jia et al.\ to conduct Type II attack~\cite{JLG22}.
Concretely, we use truck as the reference image for CIFAR10, STL10, and CIFAR100, and SGD as the optimizer.
The batch size is 32 and the learning rate is 0.001.
The $\lambda_1$ is 1 and $\lambda_2$ is 1.

\mypara{Defense Methods Implementation Details}
We utilize the source code of Neural Cleanse and STRIP and the spectral signature implementation from ART~\cite{ART} to detect backdoored models and poisoned samples.
For Neural Cleanse, we regard the downstream models as detect targets and adopt clean test sets to reverse the triggers.
STRIP is an input-level defense that detects whether the incoming input is poisoned.
In the implementation, we randomly perturb 4\% test samples, i.e., 400 samples in CIFAR10, by other 2\% samples to calculate the entropy score.
Spectral signatures defend poisoning-based backdoor attacks at the dataset level.
To fit the real usage scenario of spectral signatures, we utilize the pre-trained dataset in Type I attack and downstream datasets in Type III attack to calculate the backdoor and clean score.

\mypara{Runtime Configuration}
We perform experiments on 4 NVIDIA A100 GPUs, each of which has 40GB memory.

\begin{table}[!t]
\centering
\caption{Comparision of different optimizers.}
\label{table:optimizer}
\scalebox{0.8}{
\begin{tabular}{l|ccc}
\toprule
\multicolumn{1}{c|}{\multirow{2}{*}{Optimizer}} & \multicolumn{3}{c}{CA}                                                                 \\
\multicolumn{1}{c|}{}                           & \multicolumn{1}{c}{CIFAR10} & \multicolumn{1}{c}{CIFAR100} & \multicolumn{1}{c}{STL10} \\
\midrule
AdamW                                           & 87.71                       & 68.22                        & 95.09                     \\
SGD                                             & 86.81                       & 64.94                        & 94.71                     \\
LARS                                            & 65.15                       & 28.31                        & 46.03                     \\
\bottomrule
\end{tabular}
}
\end{table}

\subsection{Evaluation Metrics}
\label{section:eval_metrics}

Formally, the definitions for evaluation metrics are as follows:

\begin{itemize}
\item \textbf{Clean Accuracy:} The clean accuracy is the classification accuracy of a clean downstream model on the clean testing images.
\item \textbf{Test Accuracy:} The test accuracy is the classification accuracy of a backdoored downstream model on the clean testing images.
If the test accuracy of a backdoored downstream classifier is similar to the clean accuracy, the backdoor attack preserves accuracy for the downstream task.
\item \textbf{Attack Success Rate (ASR):} The ASR is the fraction of trigger-injected images that are predicted as the target class by the backdoored downstream classifier.
\item \textbf{Attack Success Rate-Baseline (ASR-Baseline):} As a baseline, ASR-Baseline is the fraction of trigger-injected images that are predicted as the target class by the clean downstream model.
\end{itemize}

\subsection{Ablation Study}
\label{section:ablation_study_appendix}

We conduct a series of ablation studies to understand the impacts of important backdoor attack components in each supply chain phase.
We summarized the results by order of the attack types.
The main findings have been reported at~\autoref{section:ablation_study_main}.

\subsubsection{Type I Attack}

\begin{figure*}[!htp]
\centering
\includegraphics[width=0.8\linewidth]{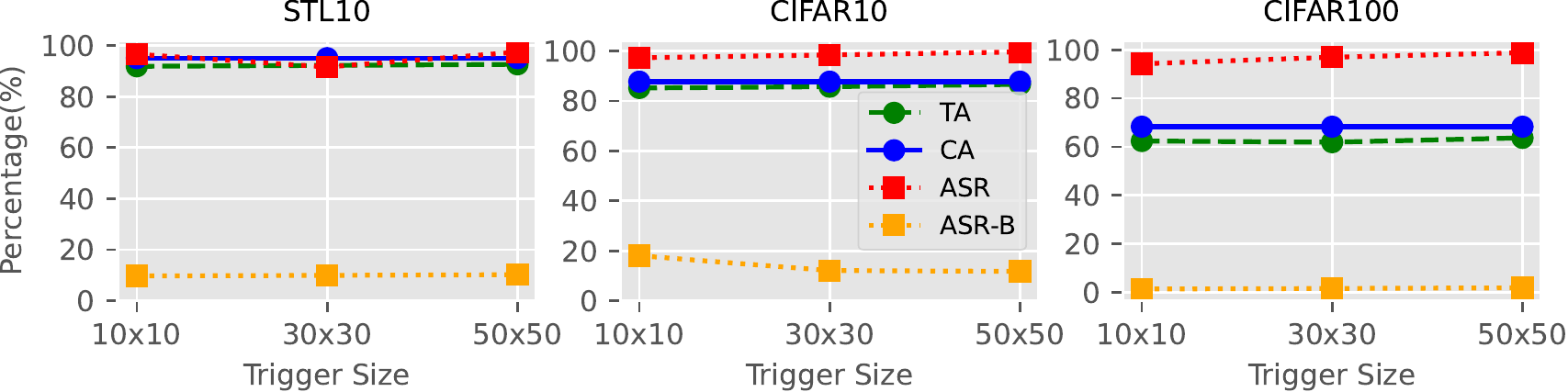}
\caption{Impacts of trigger size in Type I attack. Poisoning rate is 50\%.}
\label{figure:att_1_trigger_size}
\end{figure*}

\mypara{Impacts of Trigger Size}
\autoref{figure:att_1_trigger_size} shows the impacts of the trigger size in Type I attack. 
For all three downstream datasets, we observe that the Type I attack remains high attack success rate when the trigger is tiny. 
For instance, when the trigger size is $10\times 10$, which only occupies 0.20\% area of the whole image, the ASR can still reach 96.58\% on STL10 dataset.
The second thing we observed is that when the trigger gets larger, the ASR  increases and the test accuracy decreases.
This observation meets our expectations and results from previous work~\cite{JLG22}, as the model is more likely to notice the trigger when it becomes larger, it is naturally easier to map images with trigger to the target label.

\mypara{Impacts of Poisoning Rate}
Have already been discussed in~\autoref{section:ablation_study_main}.

\begin{figure*}[!htp]
\centering
\includegraphics[width=0.8\linewidth]{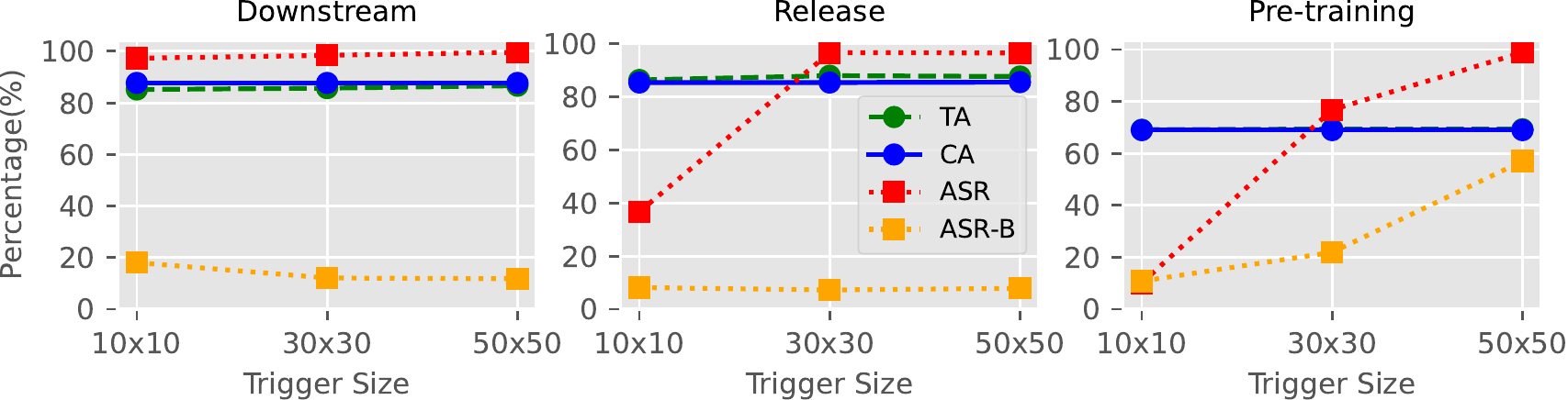}
\caption{Impact of trigger size in different model supply chain phases on CIFAR10.}
\label{figure:trigger_size_cifar10}
\end{figure*}

\subsubsection{Type II Attack}

Since Type II attacker has no capabilities to tamper with the pre-training dataset and downstream dataset (see~\autoref{table:attack_taxonomy}), poisoning rate is not a hyperparameter in Type II attack.
Here, we mainly investigate the impacts of trigger size and mask.

\begin{figure*}[!htp]
\centering
\includegraphics[width=0.8\linewidth]{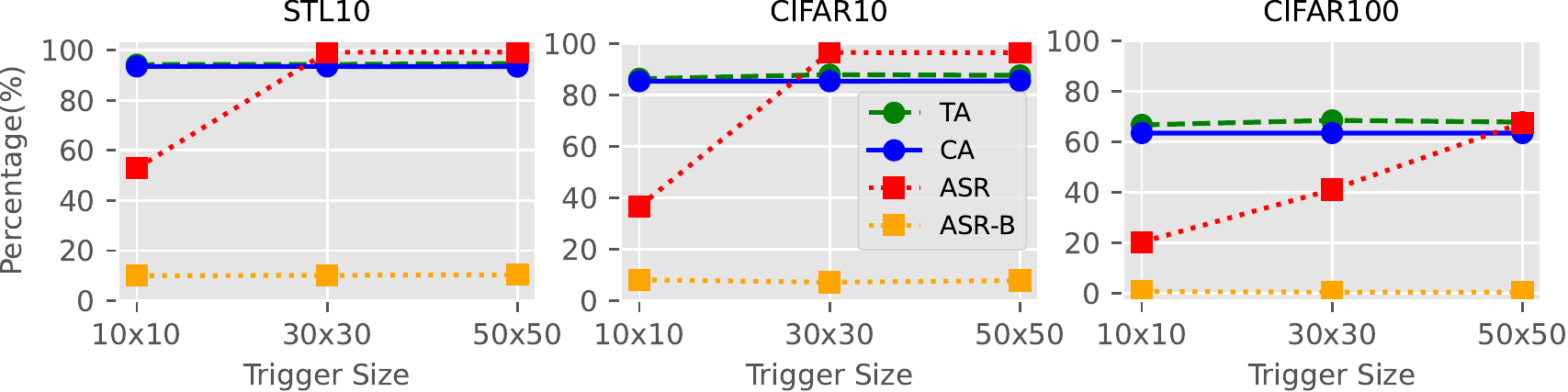}
\caption{Impacts of trigger size in Type II attack.}
\label{figure:att_2_trigger_size}
\end{figure*}

\mypara{Impacts of Trigger Size}
\autoref{figure:att_2_trigger_size} shows the impacts of trigger size in Type II attack.
Following previous observation, when trigger size enlarges, ASR increase.
For instance, the ASR on CIFAR10 increase from 36.71\% to 96.48\% when trigger size enlarges from $10\times 10$ to $50\times 50$.
However, unlike Type I attack, we do not observe a significant decrease in utility performance as the trigger becomes larger.
Take CIFAR10 as an example.
The test accuracy is 86.32\%,  87.92\%, 87.62\% on trigger $10\times 10$, $30\times 30$, and $50\times 50$, respectively.
This might be due to the attack mechanism of Type II attack.

\mypara{Impacts of Mask}
Have already been discussed in~\autoref{section:ablation_study_main}.

\begin{figure*}[!htp]
\centering
\includegraphics[width=0.8\linewidth]{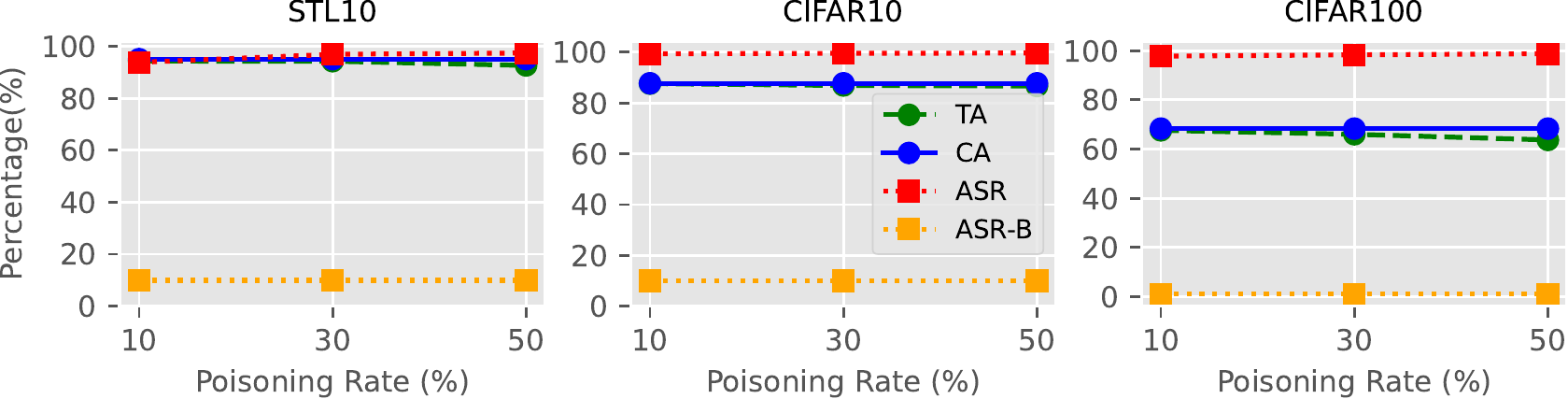}
\caption{Impacts of poisoning rate in Type I attack. Trigger size is $50\times 50$.}
\label{figure:att_1_poison_rate_tri50}
\end{figure*}

\subsubsection{Type III Attack}

Type III attack settles at the beginning of the supply chain.
Here, Type III attack refers to the general method that only poison the subset of target label.
Ablation study of trigger size and poisoning rate is based on Type III-M attack.
Then, we discuss the impacts of trigger position and trigger number, which include both Type III-R attack and Type III-M attack.
Note that since our target model is trained on ImageNet20, it does not cover all classes of CIFAR100, which means even a clean model cannot achieve good clean accuracy.
Therefore, when doing an ablation study, we replace CIFAR100 with ImageNet20 as the third dataset.
Since MIM~\cite{HCXLDG21, CDWXMWHLZW22} also uses pre-training set as the downstream training set, we believe this replacement is valuable.

The impacts of trigger number, location, and pattern have been discussed in~\autoref{section:ablation_study_main}.

\begin{figure*}[!htp]
\centering
\includegraphics[width=0.8\linewidth]{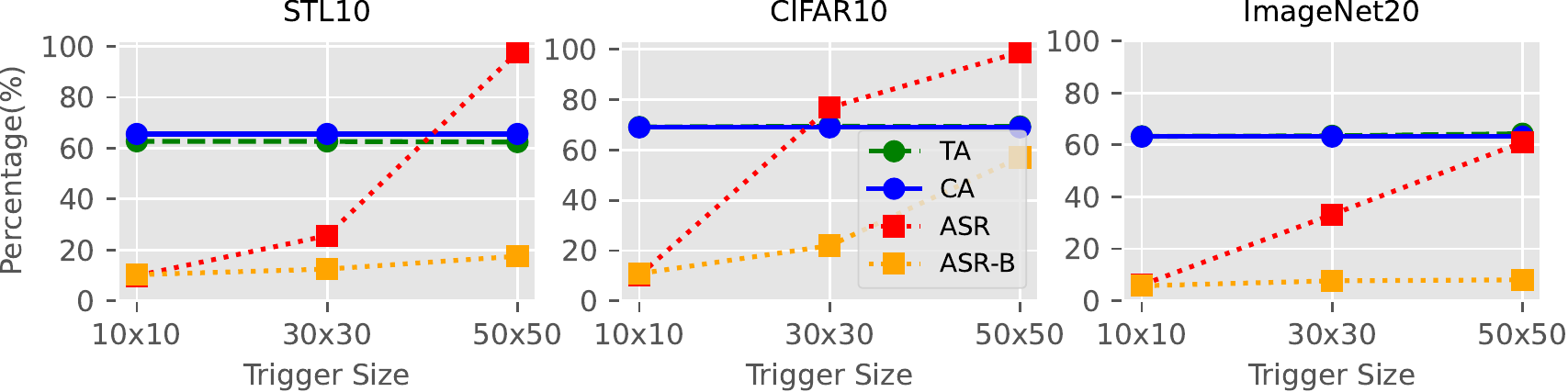}
\caption{Impacts of trigger size in Type III-M attack. Poisoning rate is 4.5\%.}
\label{figure:att_3_trigger_size}
\end{figure*}

\mypara{Impacts of Trigger Size}
\autoref{figure:att_3_trigger_size} shows the impacts of trigger size in Type III-M attack.
Similar to observation on Type I attack and Type II attack, ASR increases when the trigger becomes larger.
However, the trigger required to obtain a higher ASR is larger compared to the other two attacks.
For instance, when the trigger is $30\times 30$, ASR is 25.63\% on STL10.
And when it enlarges to $50\times 50$, the ASR rises to 97.74\%.
Besides, we also do not observe significant drops in the test accuracy as the trigger starts to expand.
Still take STL10 as an example, the test accuracy is 62.7\%, 62.64\%, and 62.39\% on Trigger $10\times 10$, $30\times 30$, and $50\times 50$, respectively.

\begin{figure*}[!htp]
\centering
\includegraphics[width=0.8\linewidth]{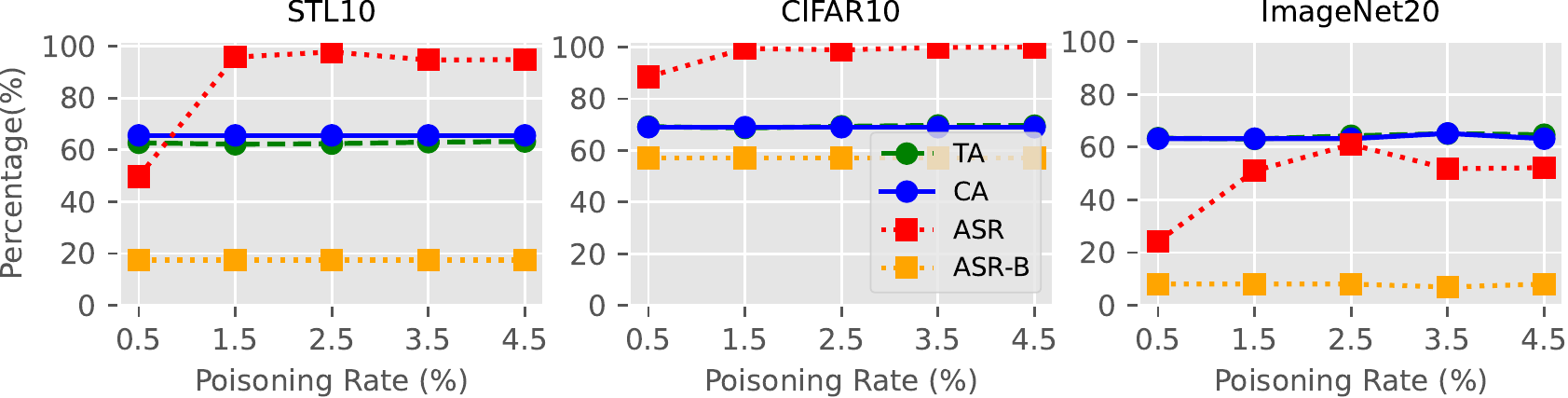}
\caption{Impacts of poisoning rate in Type III-M attack. Trigger size is $50\times 50$.}
\label{figure:att_3_poison_rate}
\end{figure*}

\mypara{Impacts of Poisoning Rate}
\autoref{figure:att_3_poison_rate} presents the impacts of poisoning rate in Type III-M attack.
Here, the poisoning rate is the rate at that the adversary poisons images of the whole dataset.
For instance, when the poisoning rate is 2.5\%, it means 50\% images of the target class, i.e., airplane, are poisoned.
We observe that when the poisoning rate reaches 1.5\%, Type III-M attack can already achieve high ASR.
For instance, the ASR is 95.76\% when the poisoning rate is 1.5\% on STL10.

\subsection{Feature Space Visualization}

\begin{figure*}[!htp]
\centering
\begin{subfigure}{0.33\linewidth}
\includegraphics[width=\linewidth]{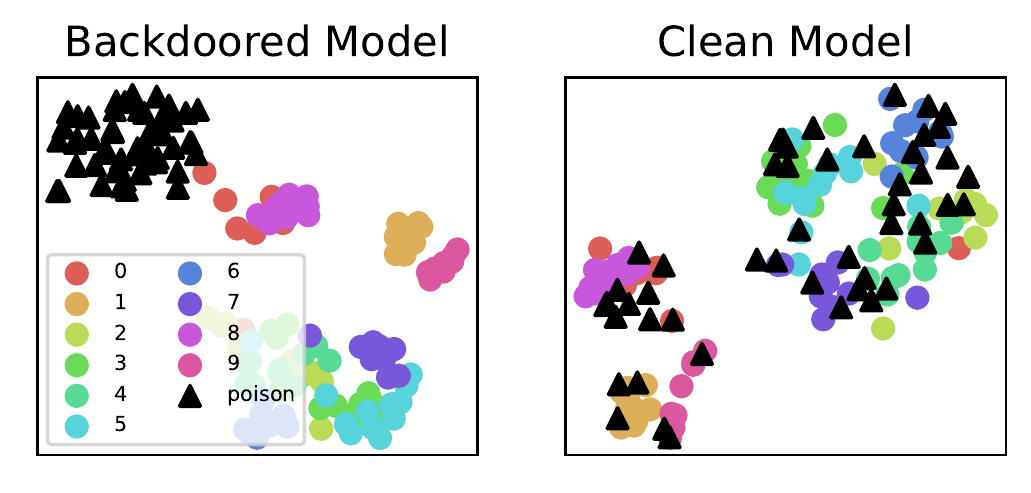}
\caption{Type I attack}
\end{subfigure}
\begin{subfigure}{0.33\linewidth}
\includegraphics[width=\linewidth]{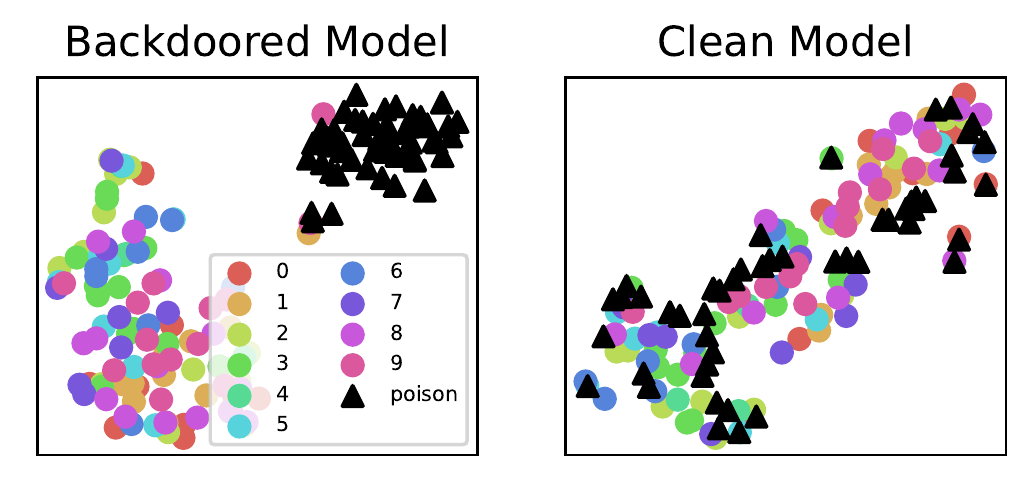}
\caption{Type II attack}
\end{subfigure}
\begin{subfigure}{0.33\linewidth}
\includegraphics[width=\linewidth]{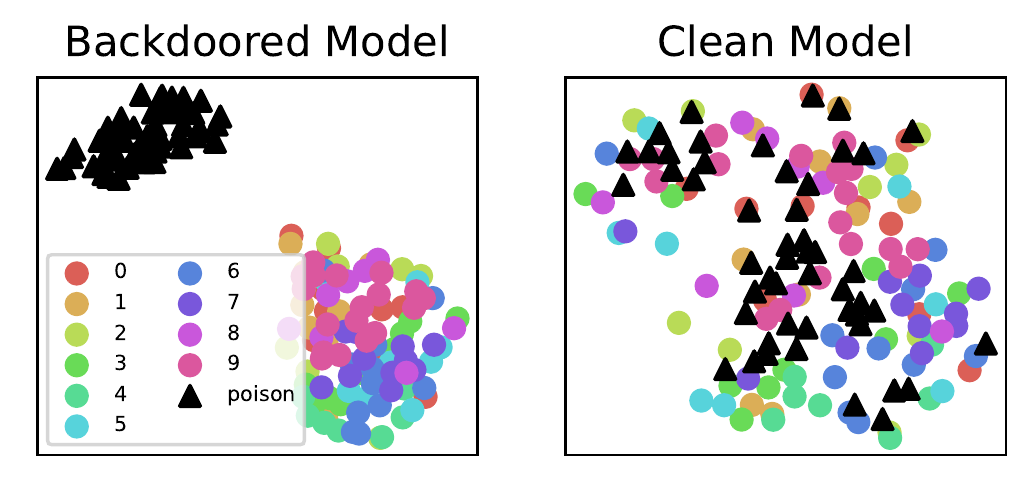}
\caption{Type III-M attack}
\end{subfigure}
\caption{t-SNE plots of feature space.}
\label{figure:tsne}
\end{figure*}

To further investigate whether the backdoor is successfully injected into the model, we visualize the feature space via t-SNE, as~\autoref{figure:tsne} shows.
We observe that for all three attacks, the poisoned samples tend to cluster together and close to the target class.
This result is highly correlated with the effectiveness of the attack.

\subsection{ImageNet20}

\autoref{table:imagenet20} shows the list of classes from ImageNet20.
All classes are randomly sampled from the class list of the original ImageNet-1k dataset~\cite{DDSLLF09}.

\begin{table}[!t]
\centering
\caption{List of classes from ImageNet20.}
\label{table:imagenet20}
\scalebox{0.8}{
\begin{tabular}{ll|ll}
\toprule
ID          & Label             & ID          & Label       \\
\midrule
n02123394   & Persian cat       & n03661043   & Library     \\
n02085936   & Maltese dog       & n07718472   & Cucumber    \\
n02489166   & Proboscis monkey  & n07734744   & Mushroom    \\
n02690373   & Airliner          & n03764736   & Milk can    \\
n03095699   & Container ship    & n03291819   & Envelope    \\
n04285008   & Sports car        & n03770439   & Miniskirt   \\
n04461696   & Tow truck         & n03124170   & Cowboy hat  \\
n01833805   & Hummingbird       & n03916031   & Perfume     \\
n01644900   & Tailed frog       & n03938244   & Pillow      \\
n03063689   & Coffeepot         & n07614500   & Ice cream   \\
\bottomrule
\end{tabular}
}
\end{table}

\begin{figure*}[!htp]
\centering
\begin{subfigure}{0.25\linewidth}
\includegraphics[width=\columnwidth]{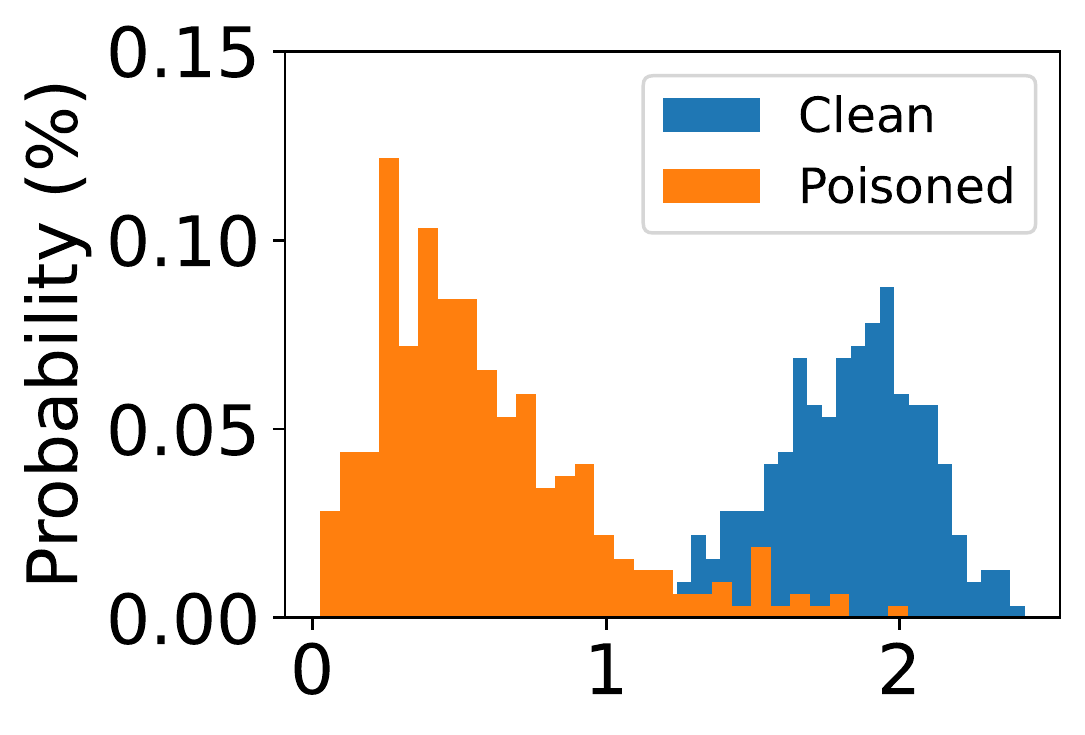}
\caption{Type I attack}
\end{subfigure}
\begin{subfigure}{0.25\linewidth}
\includegraphics[width=\columnwidth]{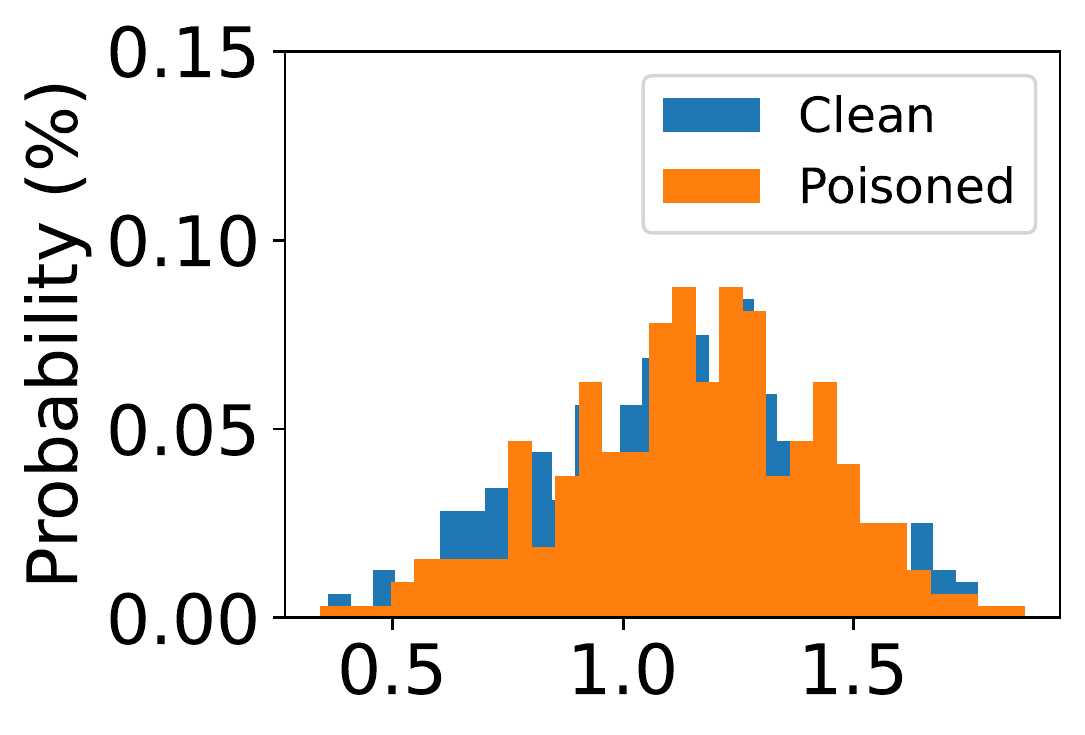}
\caption{Type II attack}
\end{subfigure}
\begin{subfigure}{0.25\linewidth}
\includegraphics[width=\columnwidth]{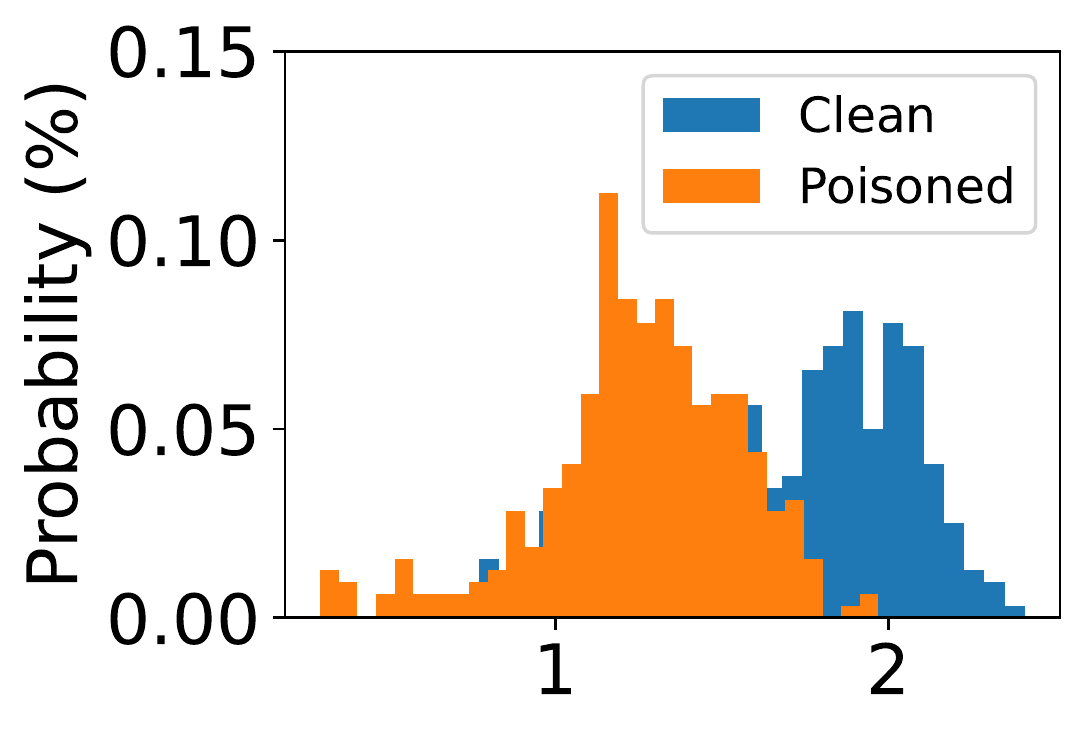}
\caption{Type III-M attack}
\end{subfigure}
\caption{Entropy distribution of STL10, calculated by STRIP.}
\label{figure:strip_stl10}
\end{figure*}

\begin{figure*}[!htp]
\centering
\includegraphics[width=0.8\linewidth]{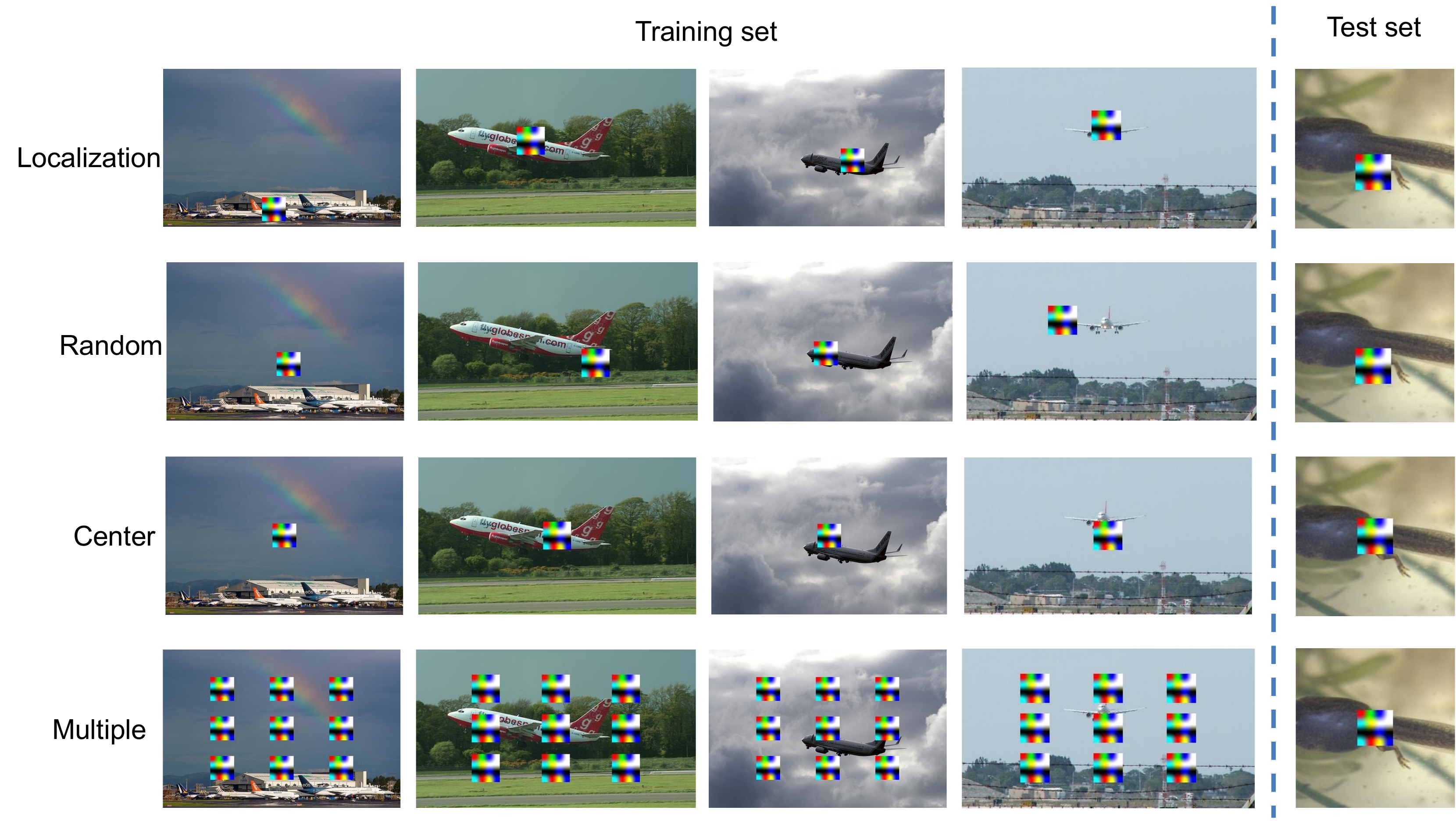}
\caption{Trigger put methods. Trigger size is $50\times 50$. For the localization method, we utilize a YOLOv5 model~\cite{yolov5} to put the trigger on the center of the target object.}
\label{figure:trigger_location}
\end{figure*}

\end{document}